\input epsf

\documentstyle[sprocl]{article}

\def\be{\begin{equation}}
\def\ee{\end{equation}}
\def\bea{\begin{eqnarray}}
\def\eea{\end{eqnarray}}

\begin{document}
\title{SELF-ORGANIZED CRITICALITY}
\author{MICHAEL CREUTZ}
\address{Physics Department, Brookhaven National Laboratory,
Upton, NY 11973, USA}
\maketitle\abstracts{ 
I review the concept of self-organized criticality, wherein
dissipative systems naturally drive themselves to a critical state
with important phenomena occurring over a wide range of length and
time scales.  Several exact results are demonstrated for the Abelian
sandpile. }

Self-organized criticality concerns a class of dynamical systems which
naturally drive themselves to a state where interesting physics occurs
on all scales\cite{btw}.  The idea provides a possible ``explanation''
of the omnipresent multi-scale structures throughout the natural
world, ranging from the fractal structure of mountains, to the power
law spectra of earthquake sizes
\cite{mcbak}.  Recent applications include such diverse topics as
evolution\cite{pmb} and traffic flow \cite{mayanagel}.  The concept
has even been invoked to explain the unpredictable nature of economic
systems, i.e. why you can't beat the stock market \cite{sorin}.

The prototypical example is a sandpile.  On slowly adding grains of
sand to an empty table, a pile will grow until its slope becomes
critical and avalanches spill over the sides.  If the slope becomes
too large, a large catastrophic avalanche is likely, and the slope
will reduce.  If the slope is too small, then the sand will accumulate
to make the pile steeper.  Ultimately one should obtain avalanches of
all sizes, with the prediction for the next being impossible without
actually running the experiment.

Self-organized criticality nicely compliments the concept of chaos.
In the latter, dynamical systems with a few degrees of freedom, say
three or more, can display highly complex behavior, including fractal
structures.  With self-organized criticality, we start instead with
systems of many degrees of freedom, and find a few general common
features.  Another attraction of this topic is the ease with which
computer models can be implemented and the elegance of the resulting
graphics \cite{xtoys}.

The original Bak, Tang, Wiesenfeld paper \cite{btw} presented a simple
model wherein each site in a two dimensional lattice has a state
specified by a positive integer $z_i$.  This can be thought of as the
amount of sand at that location, or, in another sense, as the slope of
the sandpile at that point.  Neither of these analogies is fully
accurate, for the model has aspects of each.

\begin{figure}
\epsfxsize .8\hsize
\centerline {\epsfbox{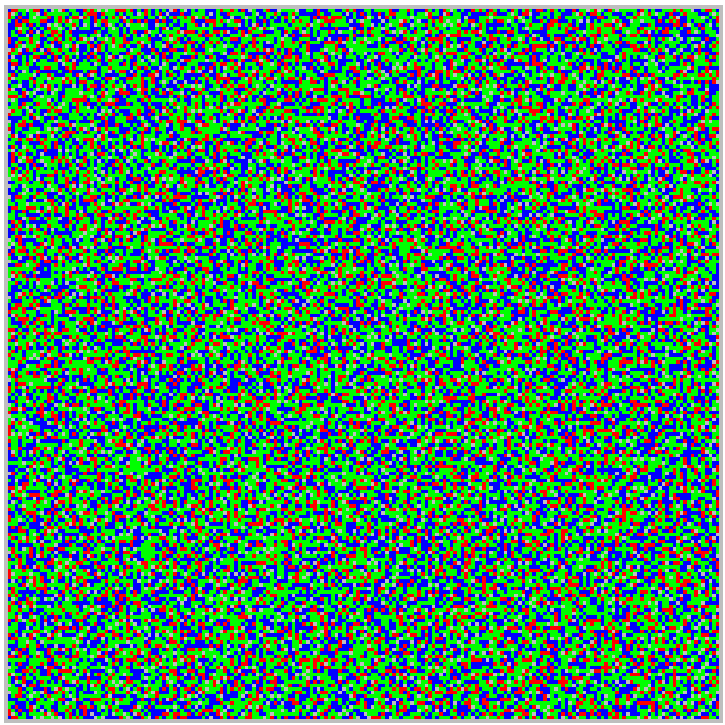}}
\caption{The sandpile model in the final stable state after adding
lots of sand to random places.  The lattice is 198 cells by 198 cells.
The color code is grey, red, blue, and green for heights 0,1,2, and 3,
respectively.  Despite the lack of obvious patterns, subtle
correlations are present; for example no two adjacent sites have
height zero.
\label{fig:sand0}}
\end{figure}

\begin{figure}
\centerline {a.
\epsfxsize .45\hsize \epsfbox{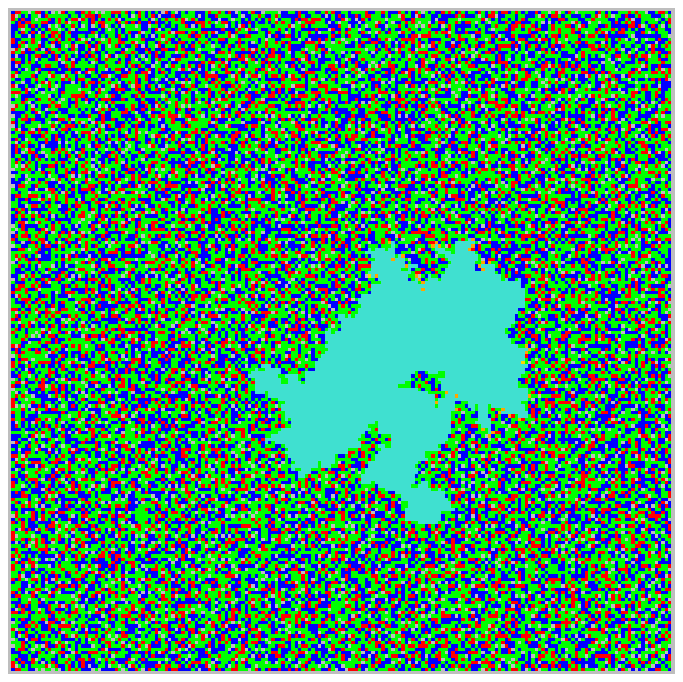}
\hskip .03\hsize 
b.
\epsfxsize .45\hsize \epsfbox{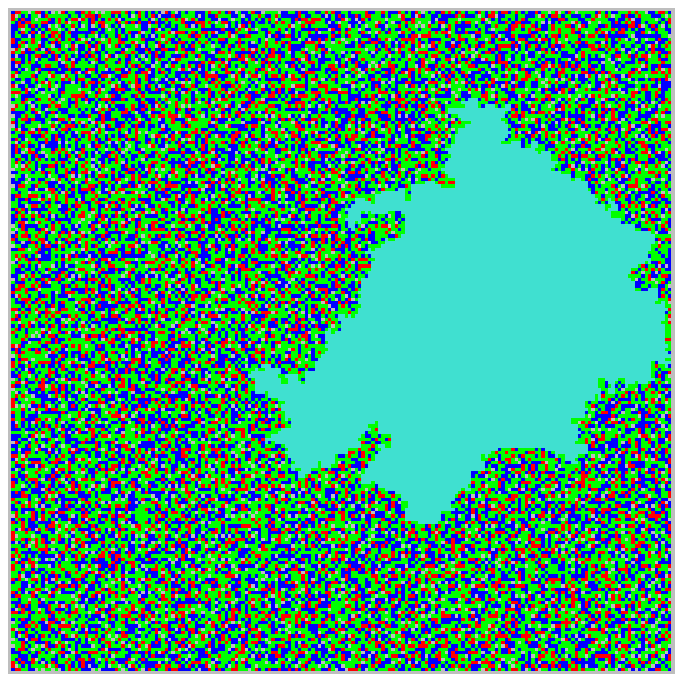}
}
\caption{An avalanche obtained by adding a small amount of sand to the
configuration in Fig.~\protect\ref{fig:sand0}.  Stable sites which
have tumbled during the avalanche are distinguished by being colored
light blue.  The still active sites on the left image are colored
yellowish brown.  The image on the right is the final state after the
avalanche has ended.  Note that the final avalanche region is simply
connected.  This is a general result proven in the text.}
\label{fig:sand1}
\end{figure}

The dynamics follows by setting a threshold $z_T$ above which any
given $z_i$ is unstable.  Without loss of generality, I take this
threshold to be $z_T=3$.  Time now proceeds in discrete steps.  In one
such step each unstable site ``tumbles'' or ``topples,'' dropping by
four and adding one grain to each of its four nearest neighbors.  This
may produce other unstable sites, and thus an avalanche can ensue for
further time steps until all sites are stable.  Fig.~\ref{fig:sand0}
shows a typical configuration on a 198 by 198 lattice after lots of
random sand addition followed by relaxation.  Fig.~\ref{fig:sand1}
shows an avalanche proceeding on this lattice.

A natural experiment consists of adding a grain of sand to a random
site and measuring the number of topplings and the number of time
steps for the resulting avalanche.  Repeating this many times to gain
statistics, the distribution of avalanche sizes and lengths displays a
power law behavior, with all sizes appearing.  In
Ref.~\cite{christensen} such experiments showed that the distribution
of the number of tumbling events $s$ in an avalanche empirically
scales as
$$
P(s)\sim s^{-1.07}
\eqno(1)
$$
and the number of time steps $\tau$ for avalanches scales as
$$
P(\tau)\sim \tau^{-1.14}
\eqno(2)
$$
This model has been extensively studied analytically.  While as yet
there is no exact calculation of these exponents, a lot is known.  In
particular, the critical ensemble is well characterized.  I will
return to these points later.

The extent to which laboratory experiments reproduce such power laws
is somewhat controversial.  A recent study of avalanche dynamics
\cite{rice} in
rice piles showed criticality with long-grain rice, but more
ambiguous results followed similar experiments with short-grain rice.

Another simple model mimics forest fires and has three possible states
per cell, empty, a tree, or a fire.  For the updating step, any empty
site can have a tree born with a small probability.  At the same time,
any existing fire spreads to neighboring trees leaving its own cell
empty.  The random growth of trees gives this rule a stochastic
nature.  As the system is made larger, the growth rate for the trees
should decrease to just enough to keep the fires going.

\begin{figure}
\hbox{
a.
\epsfxsize .45\hsize
\epsffile{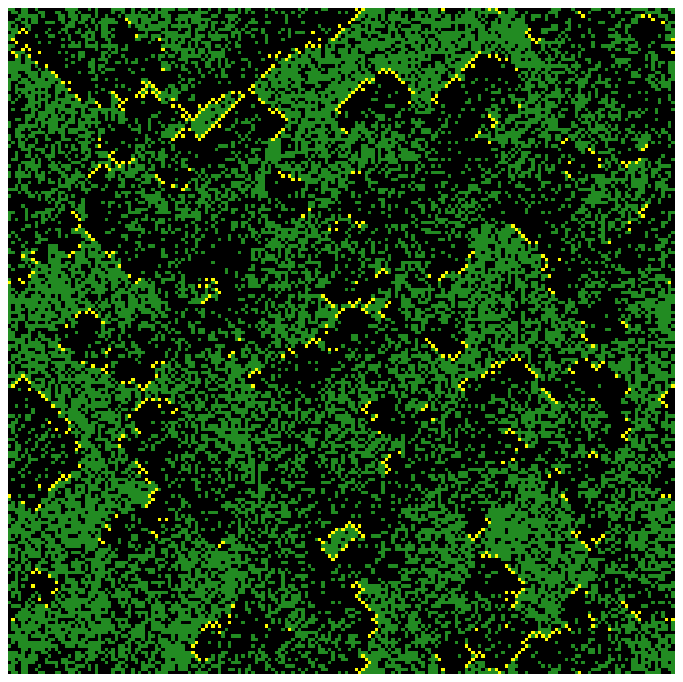}
\hskip .02\hsize
b.
\epsfxsize .45\hsize
\epsffile{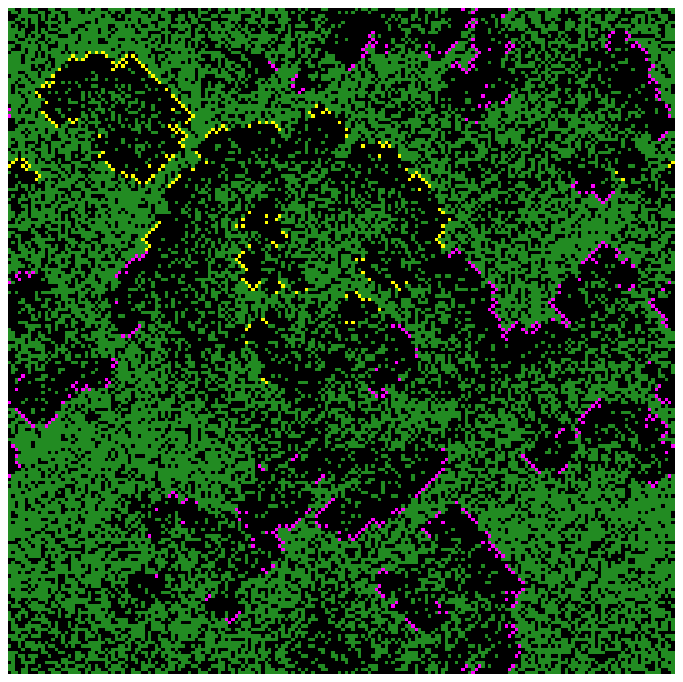}
} 
\caption{On the left is a snapshot of the forest fire model on a 198 by 198
lattice.  Trees are continuously burning at a slow rate, while fires
burn them down and spread to nearest neighbor trees.  Here the four
cell neighborhood is used.  On the right is a variation where two
species of bunnies are competing to eat a common grass.  The yellow
and purple colors here distinguish the parity of the site plus time
step.  Eventually one of the two species dominates and the other dies
out.}
\label{fig:fires}
\end{figure}

If too many trees grow, one obtains a large fire reducing their
density, while if there are too few trees, fires die out.  On a finite
system, one should light a fire somewhere to get the system started.
As the system becomes larger, the growth rate for the trees can be
reduced without the fire expiring.  In a steady state the system has
fire fronts continually passing through the system, as illustrated in
Fig.~{\ref{fig:fires}a}.  Perhaps there is a moral here to be careful
about extinguishing all fires in the real world, for this may enhance
the possibility for a catastrophic uncontrollable fire.  It is not
entirely clear whether this model is actually critical.  What seems to
happen on large systems is that stable spiral structures form and set
up a steady rotation.  For a review of this and several related
models, see Ref.~\cite{cds}.

A variation on this model has several ``species'' of fires.  Perhaps a
better metaphor is to think of different species of bunny, competing
for the same slowly growing food resource.  With the four cell
neighborhood, a natural division into species is given by the parity
of the site plus the time step.  Fig.~{\ref{fig:fires}b.}  shows a
state in the evolution of such a model when both species are present.
This situation, however, is highly unstable, with any fluctuation
favoring one species tending to grow until the competitor is
eliminated.  This model provides a discrete realization of the
``principle of competitive exclusion'' in biological systems
\cite{murray}.  Stability of a species requires that it occupy its own
niche and not compete for exactly the same resources as another.

Very little rigorous is known about general self-organized critical
systems.  However, in a series of papers, Deepak Dhar and co-workers
have shown that the sandpile model has some rather remarkable
mathematical properties\cite{dar1,dar2,dar3,dar4}. In particular, the
critical ensemble of the system has been well characterized in terms
of an Abelian group.  In the following I will generally follow the
discussion given in Refs.~\cite{mysand,mcbak}.

Dhar\cite{dar1} introduced the useful toppling matrix $\Delta_{i,j}$
with integer elements representing the change in the height, $z$ at
site $i$ resulting from a toppling at site $j$. More precisely, under
a toppling at site $j,$ the height at any site $i$ becomes $z_i -
\Delta_{i,j}$.  For the simple two dimensional sand model the toppling
matrix is thus
$$ \matrix
{
\Delta_{i,j}&= 4    &   i=j     \cr               
\Delta_{i,j} &=-1   & i, j {\rm \ \ nearest\ neighbors}     \cr 
\Delta_{i,j} &= 0   &        {\rm \ \ otherwise.}   \cr
}
\eqno (3)
$$

For this discussion there is little special to the specific lattice
geometry; indeed, the following results easily generalize to other
lattices and dimensions.  The analysis requires only that under a
toppling of a single site $i,$ that site has its slope decreased
$(\Delta_{i,i} > 0)$, the slope at any other site is either increased
or unchanged $(\Delta_{i,j} \leq 0, j \ne i)$, the total amount of
sand in the system does not increase $(\sum_j \Delta_{i,j} \ge 0)$, and,
finally, that each site be connected through toppling events to some
location where sand can be lost, such as at a boundary.

For the specific case in Eq. 3, the sum of slopes over all sites is
conserved whenever a site away from the lattice edge undergoes a
toppling. Only at the lattice boundaries can sand be lost. Thus the
details of this model depend crucially on the boundaries, which we
take to be open.  A toppling at an edge loses one grain of sand and at
a corner loses two.

The actual value of the maximum stable height $z_T$ is unimportant to
the dynamics.  This can be changed by simply adding constants to all
the $z_i$. Thus without loss of generality I consider $z_T = 3$.  With
this convention, if all $z_i$ are initially non-negative they will
remain so, and I thus restrict myself to states $C$ belonging to that
set. The states where all $z_i$ are positive and less than 4 are
called stable; a state that has any $z_i$ larger than or equal to 4 is
called unstable. One conceptually useful configuration is the
minimally stable state $C^*$ which has all the heights at the critical
value $z_T$.  By construction, any addition of sand to $C^*$ will give
an unstable state leading to a large avalanche.

I now formally define various operators acting on the states
$C$. First, the ``sand addition" operator $\alpha_i$ acting on any $C$
yields the state $\alpha_iC$ where $z_i = z_i+1$ and all other $z$ are
unchanged.  Next, the toppling operator $t_i$ transforms $C$ into the
state with heights $z^\prime_j$ where $z^\prime_j = z_j -
\Delta_{i,j}$. The operator $U$ which updates the lattice one time
step is now simply the product of $t_i$ over all sites where the slope
is unstable,
$$
UC=\prod_i t_i^{p_i}C
\eqno(4)
$$
where $p_i = 1$ if $z_i\geq 4$; $0$ otherwise.  Using $U$ repeatedly
gives the relaxation operator $R$. Applied to any state $C$ this
corresponds to repeating $U$ until no more $z_i$ change.  Neither $U$
nor $R$ have any effect on stable states.  Finally, I define the
avalanche operators $a_i$ describing the action of adding a grain of
sand to site $i$ followed by relaxation
$$
a_iC = R\alpha_iC. \eqno (5)
$$
At this point it is not entirely clear that the operator $R$ exists;
in particular, it might be that the updating procedure enters a
non-trivial cycle consisting of a never ending avalanche. 
This, however, is impossible as can be shown from the fact that
sand spreads during an avalanche.

With an edge-less system, such as under periodic boundaries, no sand
would be lost and thus cycles are expected and easily observed. These
models might be called ``Escher models" after the artist constructing
drawings of water flowing perpetually downhill and yet circulating in
the system.  While little is known about the dynamics of this
variation on the sandpile model, some studies have been done under the
nomenclature of ``chip-firing games'' \cite{chipfiring}.  A recent
paper \cite{unisand} has argued that this lossless sandpile model on
an appropriate lattice is capable of universal computation.

I now introduce the concept of recursive states. This set, denoted
${\cal R}$, includes those stable states which can be reached from any
stable state by some addition of sand followed by relaxation.  This
set is not empty because it contains at least the minimally stable
state $C^*$.  Indeed, that state can be obtained from any other by
carefully adding just enough sand to each site to make $z_i$ equal to
three.  Thus, one might alternatively define ${\cal R}$ as the set of
states which can be obtained from $C^*$ by acting with some product of
the operators $a_i$.

It is easily shown that there exist non-recursive, transient states;
for instance, no recursive state can have two adjacent heights both
being zero.  If you try to tumble one site to zero height, then it
drops a grain of sand on its neighbors.  If you then tumble a neighbor
to zero, it dumps a grain back on the original site.  One can also
show that the self-organized critical ensemble, reached under random
addition of sand to the system, has equal probability for each state
in the recursive set.  This is a consequence of the Abelian nature
of this system discussed below.

The crucial results of Refs.~\cite{dar1,dar2,dar3,dar4} are that the
operators $a_i$ acting on stable states commute.
Furthermore, when restricted to recursive states these operators
are invertable.  Thus
they generate an
Abelian group.
An intuitive argument that sand addition may be commutative uses an
analogy with combining many digit numbers under long addition.  The
tumbling operation is much like carrying, except rather than to the
next digit the overflow spreads to several neighbors.  As addition is
known to be Abelian, despite the confusing elementary-school rules, I
might expect the sandpile addition rule also to be.

These results have several amusing consequences.  One is a
determination of the number of states in the recursive set.
Without going into details, the result is
the absolute value of the determinant of the toppling
matrix $\Delta$.  For large lattices this determinant can be found
easily by Fourier transform. In particular, whereas there are $4N$
stable states, there are only
$$
   \exp\left( N\int_{(-\pi,-\pi)}^ {(\pi,\pi)} {d^2q\over (2\pi)^2}  
   \ln(4-2q_x-2q_y)\right)     \simeq (3.2102\ldots)^N     
   \eqno (6)           
$$
recursive states.  Thus starting from an arbitrary state and adding
sand, the system ``self-organizes" into an exponentially small subset
of states forming the attractor of the dynamics.
 
Following Ref.~\cite{mysand}, I now stack sand piles on top of one
another. Given stable configurations $C$ and $C^\prime$ with
configurations $z_i$ and $z_i^\prime$, I define the state $C\oplus
C^\prime$ to be that obtained by relaxing the configuration with
heights $z_i + z_i^\prime$.  Clearly, if either $C$ or $C^\prime$ are
recursive states, so is $C\oplus C^\prime$.
 
Under $\oplus$ the recursive states form an Abelian group isomorphic
to the algebra generated by the $a_i$. The addition of a state $C$
with heights $z_i$ is equivalent to operating with a product of $a_i$
raised to $z_i$, that is
$$
B \oplus  C = \left(\prod a_i^{z_i}\right)B. \eqno (7)
$$
The operation $\oplus$ is associative and Abelian because the
operators $a_i$ are.

Since any element of a group raised to the order of the group gives
the identity, it follows that $a_i^{|\Delta|} = E$.  This implies the
simple formula $a_i^{-1}= a_i^{|\Delta|-1}$.  The analog of this for
the states is the existence of an inverse state, $-C$
$$
-C = (|\Delta|-1) \otimes C. \eqno (8)
$$
Here, $n \otimes C$ means adding $n$ copies of $C$ and relaxing.  The
state $-C$ has the property that for any state $B\oplus C\oplus(-C) =
B$.

The state $I = C\oplus(-C)$ represents the identity and has the
property $I \oplus B = B$ for every recursive state $B$.  The state
which is isomorphic to the operator $a_i$ is simply $a_iI$. The
identity state provides a simple way to check if a state, obtained for
instance by a computer simulation, has reached the attractor, i.e. if
a given state is a recursive state: A stable state is in ${\cal R}$ if
and only if $C\oplus I = C$.

\begin{figure}
\epsfxsize .55\hsize
\centerline {\epsfbox{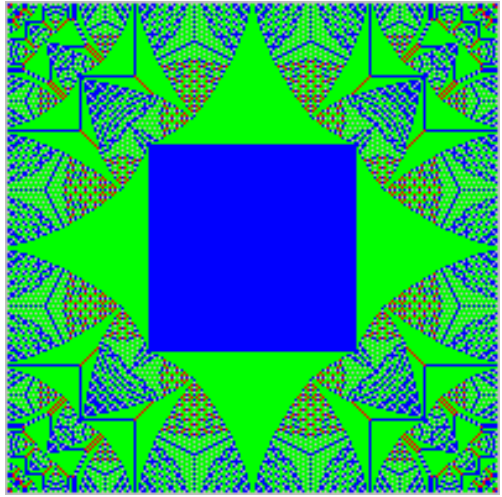}}
\caption{The identity state for the sandpile model on a 198 by 198 lattice.
The color code is grey, red, blue, and green for heights 0,1,2, and 3,
respectively.  
}
\label{fig:identity}
\end{figure}

The identity state can be constructed by taking any recursive state,
say $C^*$ and repeatedly adding it to itself to use $|\Delta| \otimes
C = I$.  However, on any but the smallest lattices, $|\Delta|$ is a
very large integer.  A more economical scheme is given in
Ref.~\cite{mysand}.  Fig.~\ref{fig:identity} shows the identity state
on a 198 by 198 lattice.  Note the fractal structure, with features on
many length scales.

\begin{figure}
\centerline {\epsfxsize .55\hsize \epsfbox{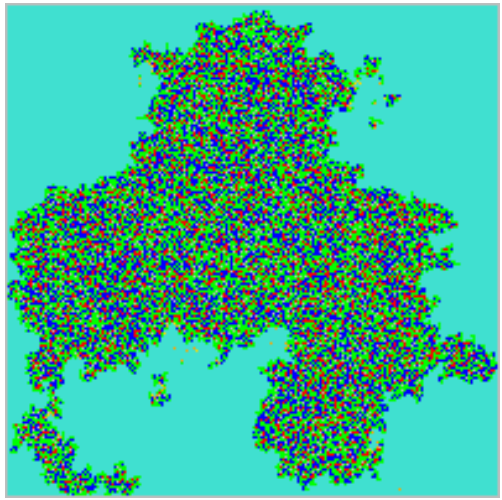}
}
\caption{
The burning algorithm being applied to the state in
Fig.~\ref{fig:sand0}.  Burnt sites are cyan, burning sites are orange,
and the remaining sites are colored as previously.  This avalanche
eventually tumbles every site exactly once.}
\label{fig:burning}
\end{figure}

Majumdar and Dhar \cite{dar4} have constructed a simple ``burning''
algorithm to determine if a state belongs to the recursive set.  For a
given configuration, first add one particle to each of the edge sites
and two particles to the corners.  This corresponds to a large source
of sand just outside the boundaries, which then tumbles one step onto
the system.  Then return to open boundaries and update according to
the usual rules.  If and only if the original state is recursive, this
will generate an avalanche under which each site of the system tumbles
exactly once.  Also, the final state after the avalanche will be
identical to the original.  However, if the state is not recursive,
some untumbled sites will remain.  Fig.~\ref{fig:burning} shows such a
process underway on the configuration of Fig.~\ref{fig:sand0}.  Here
sites which have already burned are shown in cyan, while the remaining
sites in the center have not yet tumbled.  The small number of sites
shown in orange are the still active sites, which eventually burn the
entire remaining lattice.

The burning algorithm provides a simple way to prove that the
avalanche regions are simply connected once one is in the critical
state.  In a burning process, any sub-lattice of the original will
have all of its sites tumbled onto from outside.  This is the
condition for starting a burning on the sub-lattice.  Thus, if a
configuration is in the critical ensemble for the whole lattice, then
any extracted piece of this configuration on a subset of the original
lattice is also in the critical ensemble of the extracted part.  Now
suppose that one constructs an avalanche with any initial addition to
a state from the critical ensemble.  In any subregion enclosed by this
avalanche, sand will fall from the tumbling sites on its outside.
Since the sub-lattice is itself in its own critical ensemble, this
must induce an avalanche which, by the burning algorithm, will tumble
all enclosed sites.  Thus any avalanche on a state from the critical
ensemble cannot leave untumbled any sites in a region isolated from
the boundary, i.e. an untumbled island.  This result that avalanches
must be simply connected does not follow for states outside the
recursive set, as can be easily demonstrated by considering a sandpile
with a hole of empty sites in the middle.

To conclude, simple cellular automaton models provide a rich area for
the study of complex phenomena, and in particular for systems which
self organize with physics at many scales.  I have only touched on a
few issues here, leaving out many related topics such as lattice
gasses, driven interfaces in random media, growth processes, and
evolution.  As the ease of programming and the speed of modern
computers continue to rush forward, so will the fascination with such
systems.

\section*{Acknowledgments} 
This manuscript has been authored under contract number
DE-AC02-76CH00016 with the U.S.~Department of Energy.  Accordingly,
the U.S.~Government retains a non-exclusive, royalty-free license to
publish or reproduce the published form of this contribution, or allow
others to do so, for U.S.~Government purposes.

\end{document}